\documentclass[%
 reprint,
 amsmath,amssymb,
 aps,
]{revtex4-2}

\usepackage{amssymb}
\usepackage{graphicx}
\usepackage{dcolumn}
\usepackage{bm}
\usepackage{subcaption} 
\usepackage{braket} 

\begin{document}
	
	\preprint{APS/123-QED}
	
	\title{Extended Hubbard model on fractals: d-Wave superconductivity and competing pairing channels}
	\author{Robert Canyellas} 
	\email{robert.canellasnunez@ru.nl}
    \author{Mikhail I. Katsnelson} 
	\email{m.katsnelson@science.ru.nl}
    \author{Andrey Bagrov} 
	\email{andrey.bagrov@ru.nl}
	\affiliation{
		$^{1}$Institute for Molecules and Materials, Radboud University, Heyendaalseweg 135, 6525AJ Nijmegen, The Netherlands 
	}%
	
	\begin{abstract}
		Fractal structures such as the Sierpiński gasket have been 
predicted to enhance the critical temperature of $s$-wave 
superconductivity compared to regular crystals while 
maintaining macroscopic phase coherence of Cooper pairs. 
Here we extend this analysis to order parameters with 
non-trivial symmetry by studying the extended Hubbard model 
with nearest-neighbor attraction on fractal lattices. Using 
Bogoliubov-de Gennes mean-field theory, we find that the 
Sierpiński carpet dramatically alters the competition between 
pairing channels: the predominant $d$-wave superconducting dome at half filling of the square lattice becomes unstable for the carpet, while at high and low fillings extended $s$-wave pairing gets strongly enhanced.
We attribute this to geometric frustration of 
sign-changing order parameters by the fractal boundary 
structure. On the triangular Sierpiński gasket, hybrid $s+d+id$ 
states show critical temperature enhancement comparable to 
that previously observed for pure $s$-wave pairing. Our results 
demonstrate that fractal geometry acts as a selective filter 
for pairing symmetries, with the compatibility between order 
parameter structure and lattice topology determining which 
channels are stabilized or suppressed.
	\end{abstract}
	
	\maketitle
	

The experimental advances of the past decade in atomically 
precise nanostructure fabrication based on scanning tunneling 
microscopy \cite{STM}, molecular assembly \cite{mol_ass}, supramolecular templating \cite{sup_templ}, and high-energy beam lithography \cite{lithography} have made it possible to create atomic configurations qualitatively different from naturally occurring crystals. Fractal atomic lattices have attracted particular attention 
due to their unusual geometric properties: non-integer 
Hausdorff dimension, discrete scale invariance, and 
unconventional boundary structure where bulk and edge 
interpenetrate at all length scales \cite{Gefen1,Gefen2,Gefen3,Askar_spectra_1,Askar_spectra_2,Askar_spectra_3}.

Following extensive theoretical exploration of single-particle 
quantum mechanics on fractals \cite{Kadanoff,Laskin,quantum_walks,vanVeen,Fabio,Askar_spectra_2,Askar_spectra_3,fractal_bands,fractal_light,fractal_HOT,Neupert_topology,fractal_topology,Canyellas,topological_random_fractals,fractal_conductance,westerhout_plasmons}, researchers have turned their attention to many-body correlated phenomena \cite{nodal_bands,Khemani,Ageev,Nielsen-1,Nielsen-2,fractal_SL}. In particular, magnetic properties of the repulsive Hubbard model on fractal lattices have been studied using auxiliary-field quantum Monte Carlo, revealing ferrimagnetic order on the honeycomb Sierpiński gasket \cite{AFQMC}. In \cite{BEC}, Bose-Einstein condensation on fractals and hyperbolic lattices have been studied. Ref.~\cite{Iliasov-sWave} analyzed $s$-wave superconductivity in the attractive Hubbard model using Bogoliubov-de Gennes  mean-field theory, finding that the triangular Sierpiński gasket hosts superconductivity with substantially elevated $T_c$ compared to the regular triangular lattice, while maintaining macroscopic phase coherence of the condensate. A key finding was that this enhancement occurs specifically for finitely 
ramified fractals (gasket) but not for infinitely ramified 
structures (carpet).

The $s$-wave enhancement naturally raises the question: how does 
fractal geometry affect superconducting order parameters with 
non-trivial spatial structure? This question is particularly 
intriguing for d-wave pairing (such as $d_{x^2-y^2}$ or $d_{xy}$), where the order parameter changes sign across different bond 
directions. Order parameters with sign-changing structure are expected to be particularly sensitive to geometric constraints. Consider $d_{x^2-y^2}$ pairing on the square lattice: the order parameter $\Delta_{ij}$ must be positive on horizontal bonds and negative on vertical bonds (or vice versa). This requires each site to participate 
in a ``cross'' configuration with four bonds (two positive and 
two negative) emerging in perpendicular directions. When the 
lattice is transformed into the Sierpiński carpet by 
systematically removing sites, many of these crosses are 
broken: bonds terminate at missing sites, or connect to 
regions where the local environment cannot support the 
required sign alternation. This creates geometric frustration -- 
an incompatibility between the preferred local symmetry of the 
order parameter and the constraints imposed by the fractal 
topology. 

In contrast, $s$-wave pairing (whether on-site or extended to 
nearest neighbors) maintains the same sign everywhere and is 
therefore immune to such frustration. The fractal geometry may 
thus act as a filter, selectively suppressing sign-changing 
pairing channels while allowing uniform-phase channels to 
survive or even amplify.

In this paper, we take the first step toward understanding how 
fractal geometry affects superconducting order parameters with 
non-trivial symmetry. We study the extended Hubbard model with 
both on-site ($U$) and nearest-neighbor ($V$) attractive 
interactions: the minimal model capable of hosting $d$-wave 
pairing. Working within the Bogoliubov-de Gennes mean-field 
framework \cite{BdG1,BdG2,BdG3}, we self-consistently solve for both the anomalous 
pairing amplitudes $\Delta_{ij}$ and charge densities $n_i$. Since the 
attractive Hubbard model does not develop magnetic order, we 
do not include spin-density channels in the mean-field 
treatment.

The paper is organized as follows. Section I introduces the 
extended Hubbard model and our computational methods: 
Bogoliubov-de Gennes mean-field theory (BdG) for moderate system 
sizes and the kernel polynomial method (KPM) \cite{KPM} for larger 
fractals. Section II presents the core of the results, starting with benchmarking our methods against the known deterministic quantum Monte Carlo results \cite{SousaJunior2024HalfFilledEHM}, proceeding with phase diagrams and critical temperatures for different fractal geometries and pairing 
symmetries, and concluding with the superfluid stiffness analysis aimed to verify macroscopic phase coherence \cite{Scalapino}. Section IV concludes with a discussion of physical mechanisms, 
connections to related work, and future directions.

	\section{Model and methods}
    \subsection{Extended Hubbard model}
    
    	The tight-binding Hamiltonian of the extended Hubbard model is
    	\begin{equation}
            \label{eq:Hamiltonian} 
    		\begin{split}
    			H &= -t\sum_{\langle i,j \rangle,\sigma} c^{\dagger}_{i,\sigma}c_{j,\sigma}-\mu\sum_{i,\sigma}n_{i,\sigma} \\
    			&+\frac{U}{2}\sum_{i,\sigma} n_{i,\sigma} n_{i,\bar{\sigma}}+\frac{V}{2}\sum_{\langle i,j \rangle,\sigma,\sigma'} n_{i,\sigma} n_{j,\sigma'}
    		\end{split}
    	\end{equation}
    	where $c^{\dagger}_{i,\sigma}$, $c_{i,\sigma}$ are the fermionic creation and annihilation operators, and $n_{i, \sigma} = c^{\dagger}_{i,\sigma}c_{i,\sigma}$ is the charge density at site $i$ with spin index $\sigma = \{\uparrow,\downarrow\}$. The hopping parameter representing the kinetic energy is given by $t$, the chemical potential $\mu$, the on-site and nearest-neighbor interactions are $U$ and $V$ respectively (throughout the paper we will be mostly considering the regime of attraction, $U<0$, $V<0$). The sum $\langle i,j \rangle$ runs over nearest-neighbor pairs. 
        The system is analyzed at a mean-field level; using Wick's theorem the interaction terms are decoupled  obtaining the Hartree (H), the Fock (F), and the anomalous expectation value terms (A): 
        \begin{equation}
         \quad c^{\dagger}_{i,\sigma}c_{i,\sigma}c^{\dagger}_{j,\sigma'}c_{j,\sigma'} \rightarrow H_{\mbox{H}} + H_{\mbox{F}} + H_{\mbox{A}},   \label{eq:HFA}
        \end{equation}
        where
        \begin{equation*}
            \begin{split}
               &H_{\mbox{H}} = \quad c^{\dagger}_{i,\sigma}c_{i,\sigma} \langle c^{\dagger}_{j,\sigma'}c_{j,\sigma'}\rangle +  \langle c^{\dagger}_{i,\sigma}c_{i,\sigma}\rangle c^{\dagger}_{j,\sigma'}c_{j,\sigma'} \\ - &\langle  c^{\dagger}_{i,\sigma}c_{i,\sigma}\rangle \langle c^{\dagger}_{j,\sigma'}c_{j,\sigma'}\rangle\\
                &\quad H_{\mbox{F}} = - c^{\dagger}_{i,\sigma}c_{j,\sigma'} \langle c^{\dagger}_{j,\sigma'}c_{i,\sigma}\rangle - \langle c^{\dagger}_{i,\sigma}c_{j,\sigma'}\rangle c^{\dagger}_{j,\sigma'}c_{i,\sigma} \\ +&\langle c^{\dagger}_{i,\sigma}c_{j,\sigma'}\rangle \langle c^{\dagger}_{j,\sigma'}c_{i,\sigma}\rangle\\
                &\quad H_{\mbox{A}} = c^{\dagger}_{i,\sigma}c^{\dagger}_{j,\sigma'} \langle c_{j,\sigma'}c_{i,\sigma}\rangle + \langle c^{\dagger}_{i,\sigma}c^{\dagger}_{j,\sigma'} \rangle  c_{j,\sigma'}c_{i,\sigma} \\ -& \langle c^{\dagger}_{i,\sigma}c^{\dagger}_{j,\sigma'} \rangle \langle c_{j,\sigma'}c_{i,\sigma}\rangle, 
            \end{split}
        \end{equation*}
        These mean-field contributions renormalize the system parameters with an effective site-dependent chemical potential and hopping amplitudes. In the absence of spin–orbit coupling and magnetic fields, spin is conserved, so spin-off-diagonal normal averages vanish, $\langle c^{\dagger}_{i,\sigma}c_{j,\sigma'}\rangle =0$ for all $i,j$. 
        We would like to note that including the Hartree term involving the nearest-neighbor interactions causes the BdG scheme to converge to a false optimum with either zero or complete filling. This can be bypassed by fixing the particle number explicitly and treating the chemical potential self-consistently. The resulting phase diagrams then match those computed without nearest-neighbor Hartree term, which we present in this paper.

        Substituting \eqref{eq:HFA} into~\eqref{eq:Hamiltonian}, we obtain the mean-field single-particle Hamiltonian given by
    	\begin{equation}
    		\label{eq:MF_H} 
    		\begin{split}
    			H_{MF} &= -\sum_{\langle i,j \rangle,\sigma} \big[t + V\langle c^{\dagger}_{j,\sigma}c_{i,\sigma}\rangle\big]c^{\dagger}_{i,\sigma}c_{j,\sigma}\\
                &\quad +\sum_{\langle i,j \rangle,
                \sigma}\big[-\mu +U\langle n_{i,\bar\sigma} \rangle + V\langle n_{j} \rangle \big]n_{i,\sigma} \\
    			&\quad +\sum_{i,j}\Delta_{ij}c^{\dagger}_{i,\uparrow}c^{\dagger}_{j,\downarrow}+\Delta^{*}_{ij}c_{j,\downarrow}c_{i,\uparrow},
    		\end{split}		
    	\end{equation}

    where we define the singlet-pairing potential as
    \begin{equation}
        \label{eq:sc_op}
        \Delta_{ij} =
        \begin{cases}
         \frac{U}{2}( \langle c_{i,\uparrow}c_{j,\downarrow} \rangle - \langle c_{i,\downarrow}c_{j,\uparrow} \rangle) & \text{if } i = j \\
         \frac{V}{2}( \langle c_{i,\uparrow}c_{j,\downarrow} \rangle - \langle c_{i,\downarrow}c_{j,\uparrow} \rangle)  & \text{if } i \ne j.
        \end{cases}
    \end{equation}
    Together with 
    \begin{equation}
        \label{eq:charge}
        \rho_{ij,\sigma} = \langle c^{\dagger}_{i,\sigma}c_{j,\sigma}\rangle,
    \end{equation}
    these equations constitute the mean-field self-consistency conditions.
    
    Due to the inhomogeneity of the fractal geometry, a Fourier transform is not useful, and the problem must be treated in real space. We therefore employ two complementary numerical approaches: self-consistent BdG calculations and the kernel polynomial method (KPM). The BdG approach provides full access to the quasiparticle spectrum and wavefunctions, but its computational cost grows rapidly with system size. To access larger fractal lattices, we use the KPM, which allows us to compute the mean fields without explicit diagonalization of the Hamiltonian.
    
    \subsection{Bogoliubov-de Gennes equations}
    The Bogoliubov–de Gennes equations are derived by performing a Bogoliubov canonical transformation:
    \begin{equation*}
		\begin{split}
		\label{eq:bdg} 
		c_{i,\sigma}=\sum_{n}^{'}(u_{i\sigma}^{n}\gamma_n-\sigma v_{i\sigma}^{n*}\gamma^{\dagger}_{n})\\
		c^{\dagger}_{i,\sigma}=\sum_{n}^{'}(u_{i\sigma}^{n*}\gamma^{\dagger}_n-\sigma v_{i\sigma}^{n}\gamma_{n}),
		\end{split}		
	\end{equation*}
    where the dashed sums run over positive-energy $E_n>0$ states in order to avoid double counting, and $\gamma^{\dagger}_{n}$ and  $\gamma_{n}$ are the corresponding quasiparticle creation and annihilation operators. In this basis, the effective Hamiltonian becomes diagonal and due to the absence of spin-mixing terms, the $4N\times4N$ BdG Hamiltonian block-diagonalizes into two independent $2N\times2N$ Hamiltonians corresponding to two equivalent Nambu sectors related by spin inversion. The problem then reduces to solving the eigenvalue equation
    \begin{equation}
        \begin{pmatrix}
            \hat{h} & \hat{\Delta} \\
            \hat{\Delta}^\dagger & -\hat{h}^{*}
        \end{pmatrix}
        \begin{pmatrix}
            u_{n\uparrow} \\
            v_{n\downarrow}
        \end{pmatrix}
            = E_{n}
        \begin{pmatrix}
            u_{n\uparrow} \\
            v_{n\downarrow}
        \end{pmatrix}
    \end{equation}
    where $\hat{h}$ includes the kinetic term, chemical potential, Hartree, and Fock shifts. We start from an initial guess for the order parameter $\hat{\Delta}^{0}$, diagonalize the effective Hamiltonian, and compute the mean fields defined in Eqs.~\eqref{eq:sc_op} and~\eqref{eq:charge}, which after the change of basis read 
	\begin{gather}
        \rho_{ij,\uparrow} =\sum_n  u^{n*}_{i\uparrow}u^{n}_{j\uparrow} f(E_n),\quad
       \rho_{ij,\downarrow} =\sum_n  v^{n*}_{i\downarrow}v^{n}_{j\downarrow} f(-E_n),\quad\\
		n_{i,\uparrow} =\sum_n \lvert u^n_{i\uparrow}\rvert^{2} f(E_n),\quad 
        n_{i,\downarrow} = \sum_n \lvert v^n_{i\downarrow}\rvert^{2} f(-E_n),\\
		\Delta^{\text{new}}_{ij} =  \frac{W_{i,j}}{4} \sum_{n}\Big[ u_{i\uparrow}^{n}v_{j\downarrow}^{n*}+u_{j\uparrow}^{n}v_{i\downarrow}^{*}  \Big] \tanh\Bigg(\frac{E_n}{2T}\Bigg),
    \end{gather}
where the sums now run over the full BdG spectrum, including both positive- and negative-energy states.

We iterate the self-consistency equations starting from an initial guess for the order parameter. At each iteration, the newly computed pairing field $\Delta^{\text{new}}$ is mixed with the previous step result according to     
$\Delta^{k+1}=\Delta^{k}-\eta (\Delta^{k}-\Delta^{\text{new}})$, where mixing parameter $\eta \in [0,1]$ can be either fixed or adaptive. Inspired by learning rate scheduling in machine learning, we employ different updates methods such as Root Mean Square propagation (RMSprop) \cite{RMSProp} and Adaprive Gradient (Adagrad) \cite{Adagrad}. 

    \subsection{Superfluid stiffness}
    A finite pairing amplitude alone is not sufficient for superconductivity; Cooper pairs must also maintain global phase coherence. This coherence can be probed through the static current response to a uniform vector potential. The corresponding quantity is the superfluid stiffness $D_{s}$~\cite{Scalapino}, which measures the rigidity of the condensate phase and is given by
    \begin{equation}
        \frac{D_{s}}{\pi} = \Pi_{xx}(q_{x}=0,q_{y}\rightarrow0, \omega=0)-\langle K_{x}\rangle,
    \end{equation}
    where $\Pi_{xx}$ is the retarded current-current correlator and $\langle K_{x}\rangle$ the kinetic energy. For the derivation of this equation, we refer to our last work where all terms are shown in the appendix \cite{Iliasov-sWave}.
    
    \subsection{Chebyshev expansion and Kernel polynomial method} 
    The idea of this method is to expand the spectral density, which is the difference between the retarded and advanced Green's functions $\hat{d}(w)=\hat{G}^{R}(\omega)-\hat{G}^{A}(\omega)$, in terms of the Chebyshev polynomials $\phi_{n}(x)$ and computed the mean-fields through this quantity. We follow~\cite{Nagai_kpm} where a more detailed description can be found. This orthogonal polynomials are defined in the real interval $x\in[-1,1]$ and therefore first the Hamiltonian needs to be rescaled to map its spectrum into this range. The bounds of the spectrum $E_{\text{max}}$ and $E_{\text{min}}$ can be estimated by a Lanczos procedure however it is enough if we define them as $E_{\text{max}}=10 t -\mu$ and $E_{\text{min}}=-10t+\mu$. The linear transformation is then given by $\tilde{H} = (H-\mathbb{I}b)/a$ with $a 
    =(E_{\text{max}} - E_{\text{min}})/2$ and $b =(E_{\text{max}} + E_{\text{min}})/2$.\\
    
    By defining the $2N$ unite component vectors $\bm{e}(i)$ and $\bm{h}(i)$ that have components $[\bm{e}(i)]_\gamma=\delta_{i,\gamma}$ and $[\bm{h}(i)]_\gamma=\delta_{i+N,\gamma}$ the projection of the rescaled Hamiltonian onto the $n+1$ Chebyshev polynomial is given by the recursive relation  
    \begin{equation}
        \bm{p}_{n+1}= 2 \tilde{H}\bm{p}_{n}-\bm{p}_{n-1} \quad (n\geq1),
    \end{equation}
    with $\bm{p}_{0}= \bm{p}$ and $\bm{p}_{1}= \tilde{H}\bm{h}$. The mean fields can be computed then by the following expression
    \begin{equation}
        \begin{split}
        \langle c^{\dagger}_{i}c_{j}\rangle&=-\frac{1}{2\pi i} \int_{-\infty}^\infty f({\omega})\bm{e}(j)^{T}  \hat{d}(\omega) \bm{e}_n(i)\\
        &=\sum_{n=0}^{\infty}\bm{e}(j)^{T}\bm{e}_n(i)\frac{\mathcal{T}_n}{w_n},
        \end{split}
    \end{equation}
    \begin{equation}
        \begin{split}
        \langle c_{i}c_{j}\rangle&=-\frac{1}{2\pi i} \int_{-\infty}^\infty f({\omega})\bm{e}(j)^{T}  \hat{d}(\omega) \bm{h}_n(i)\\
        &=\sum_{n=0}^{\infty}\bm{e}(j)^{T}\bm{h}_n(i)\frac{\mathcal{T}_n}{w_n},
        \end{split}
    \end{equation}
    where $f(\omega) = \frac{1}{e^{\beta\omega}+1}$ is the Fermi-Dirac distribution, $\omega_n=\frac{\pi}{2}(1+\delta_{n,0})$, and
    \begin{equation}
        \label{eq:Tn}
        \mathcal{T}_n=\int_{-1}^{1}dx f(ax+b)W(x)\phi_n(x)
    \end{equation}
    with $W(x)=\frac{1}{\sqrt{1+x^2}}$.  At the zero temperature limit the integrals of eq.~\ref{eq:Tn} can be solved analytically and are given by
    \begin{equation}
        \begin{split}            
        \mathcal{T}_0&=\pi - \arccos (-b/a),\\
        \mathcal{T}_{n\neq 0} &= -\frac{\sin[n \arccos(-b/a)]}{n}.
        \end{split}
    \end{equation}
    
\subsection{Fixing order parameter symmetry}
To explore different pairing symmetries within our self-consistent framework, we employ symmetry-constrained ans\"atze and tailored initial seeds for the order parameter $\Delta_{i,j}$. This protocol allows us to systematically explore how fractal geometry affects the stability of different pairing symmetries and whether certain lattice structures favor specific order parameter channels over others. For example, for the purpose of stability diagnostic of the $d$-wave pairing, we enforce $\Delta_{ii} = 0$ throughout
the self-consistent cycle, suppressing any on-site pairing amplitude, and $\Delta_{ij}$ for the nearest neighboring sites $i$ and $j$ are initialized with a sign structure reflecting the $d$-wave
nodal pattern: positive along bonds in the x-direction
and negative along bonds in the y-direction (or vice
versa). This constraint ensures that the converged solution, if stable, exhibits the characteristic sign-changing structure of $d$-wave order.

However, while the Bogoliubov–de Gennes equations can in principle admit multiple self-consistent solutions corresponding to local extrema of the mean-field free energy, on fractal lattices we rarely find stable ground states with purely $d$-wave or purely chiral $d+id$ order. Whenever an $s$-wave component is allowed, the iterations converge to a mixed state with lower free energy. Specifically:
\begin{itemize}
    \item For extended $s$-wave pairing, which differs from pure on-site $s$-wave by incorporating nearest-neighbor correlations without introducing sign changes, we allow both on-site and nearest-neighbor components, but initialize all elements with non-negative values. This permits the formation of Cooper pairs with both on-site and inter-site character while maintaining a uniform phase across all bonds.
    \item For mixed $s+d$ states, we allow both on-site and nearest-neighbor pairing amplitudes, and construct the initial seed as a superposition of $s$-wave and 
$d$-wave components. On fractal lattices, the self-consistent solution always retains a finite extended 
$s$-wave component. This indicates that neither a purely $d$-wave state nor a combination of local (on-site) $s$-wave and $d$-wave order parameters is energetically stable.
\item For $s+d+id$ states, we initialize a complex-valued order parameter in which one of the $d$-wave components acquires a relative phase with respect to the other, while the $s$-wave component is chosen to be real.
\end{itemize}  
	\section{Results}
    \subsection{Square and carpet at the neutrality point}
    Before exploring the full parameter space of the extended Hubbard model, we validate our numerical implementation and establish appropriate interaction strengths for subsequent analysis. This benchmark serves two purposes. First, we reproduce known quantum Monte Carlo results for the $U$-$V$ phase diagram of the extended Hubbard model on the square lattice at half-filling and fixed temperature $T = 0.001$ \cite{SousaJunior2024HalfFilledEHM}. This confirms that our mean-field treatment captures the essential physics despite its approximations. Second, since the remainder of this work focuses on specific values of $U$ and $V$, we must verify that these parameters lie within a regime supporting superconductivity rather than phase separation, which can emerge when both $U$ and $V$ become too strongly attractive \cite{vanLoon_2018}.

Figure 1 shows the $U$-$V$ phase diagram at half-filling for both the regular square lattice (left panel) and the Sierpiński carpet of generation $G = 2$ (right panel). For the square lattice, we recover the expected phase structure: s-wave pairing dominates at moderate on-site attraction with weak or repulsive nearest-neighbor interaction, while $d_{x^2-y^2}$ pairing becomes stable when nearest-neighbor attraction $V$ is sufficiently strong. At the strongest attraction strengths, charge density wave order appears, signaling proximity to phase separation. This phase diagram agrees qualitatively with quantum Monte Carlo calculations \cite{SousaJunior2024HalfFilledEHM}, confirming that our Bogoliubov-de Gennes mean-field approach captures the correct competition between pairing symmetries and the onset of charge ordering.

The Sierpiński carpet phase diagram (right panel of Fig. 1) reveals a striking modification of the phase structure. The $s$-wave region shrinks dramatically, while the $d$-wave phase disappears entirely, replaced by charge density wave order across much of the parameter space where $d$-wave pairing was stable on the regular lattice. This suppression of $d$-wave superconductivity can be attributed to the geometric frustration inherent in the carpet structure. As discussed in App.\ref{AppendixA}, $d$-wave pairing on the square lattice relies on ``elementary crosses'' formed by four bonds emerging from each site, with the order parameter alternating sign between perpendicular directions. The systematic removal of sites in the Sierpiński carpet construction breaks many of these crosses, frustrating the emergence of coherent d-wave order. In contrast, $s$-wave pairing, being isotropic, proves more resilient to the geometric constraints, though even it becomes less stable relative to charge ordering.

Based on these results, for the detailed $T$-$\mu$ phase diagrams presented in subsequent sections, we select $U = -1 $ and $V = -1$, which places the system within the superconducting regime on the regular lattice while remaining well away from the phase separation boundary. This choice allows us to explore how fractal geometry modifies superconducting pairing without considering the competing charge-ordered phases dominating other parts of the phase diagram.
    \begin{figure*}[t] 
        \centering
        \includegraphics[width=0.8\textwidth]{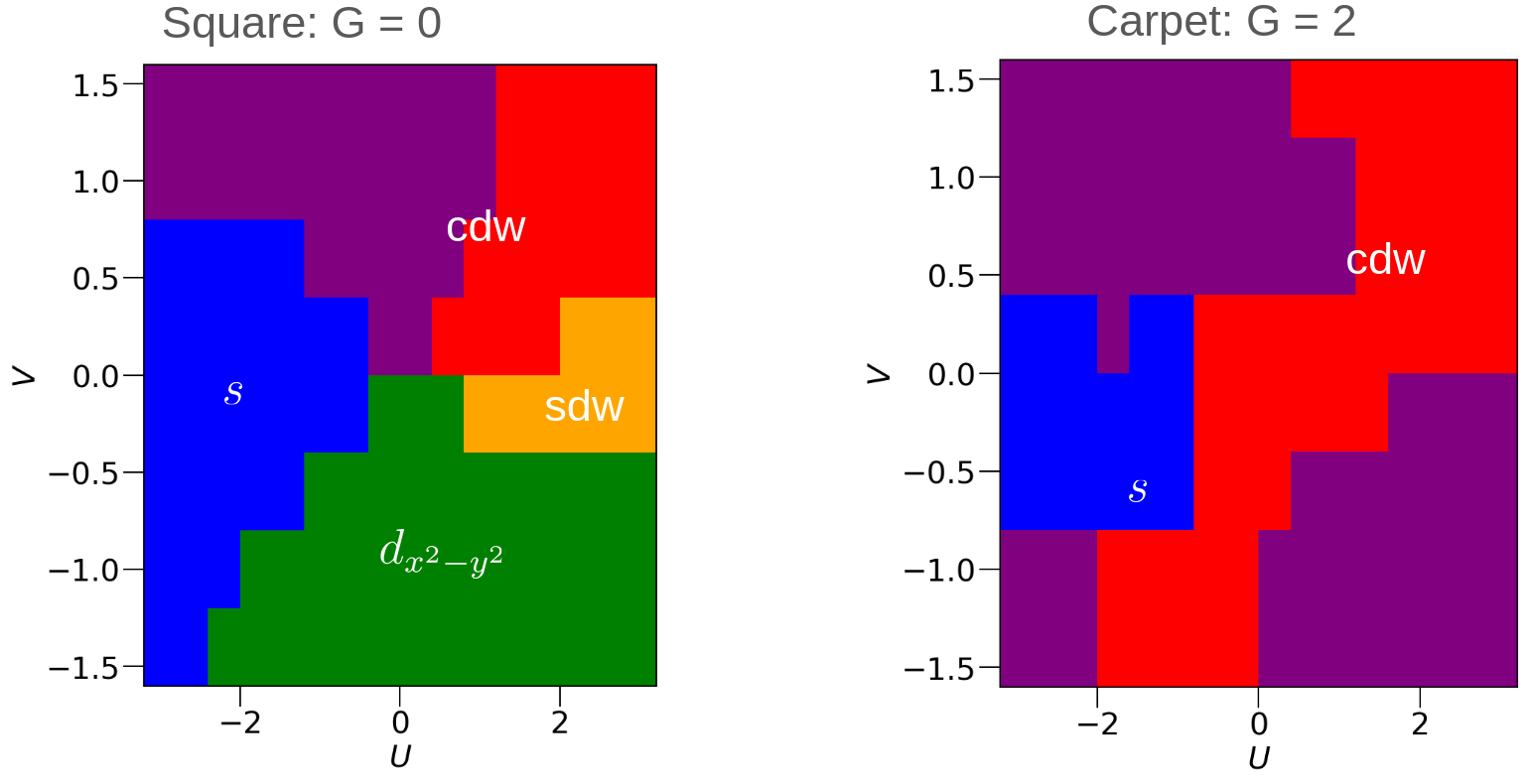}
        \caption{$U-V$ phase diagram of the extended Hubbard model on the square (left) and Sierpinski carpet (right) lattices at half-filling.}
        \label{fig:neutrality}
    \end{figure*}
    \subsection{Sierpinksi carpet}
    Having established the $U-V$ parameter regime that supports superconductivity, we now examine how fractal geometry affects the competition between different pairing symmetries. Fig. 2 shows the $T-\mu$ phase diagram for both the regular square lattice (left panel) and the Sierpiński carpet of generation $G=3$ (right panel) at fixed $U=-1$ and $V=-1$. To capture the competition between pairing channels, we allow both extended $s$-wave (on-site plus nearest-neighbor with uniform sign on a given coordination shell) and $d_{x^2-y^2}$ components in the order-parameter ansatz.

On the regular square lattice, the phase diagram exhibits a pronounced central dome dominated by $d$-wave pairing, extending to critical temperatures $T_c \simeq 0.14t$. The $d$-wave condensate reaches its maximum amplitude $\Delta \simeq 0.06$ near optimal doping. In contrast, extended $s$-wave pairing appears only in two weak domes at the edges of the superconducting region, with considerably lower critical temperatures and smaller condensate amplitudes. This hierarchy reflects the energetic favorability of $d_{x^2-y^2}$ pairing for the chosen interaction parameters on the square lattice, where the nearest-neighbor attraction $V$ promotes sign-changing order parameters that maximize pairing on perpendicular bonds.

The Sierpiński carpet phase diagram reveals a substantial restructuring of this hierarchy. The central $d$-wave dome present on the regular lattice is no longer realized as a stable pure $d$-wave solution in the corresponding parameter region. Instead, the dominant superconducting state develops a predominantly sign-definte extended-$s$ character, Fig.\ref{fig:delta} (left panel), and its critical temperature increases relative to the regular lattice case. The resulting condensate reaches amplitudes comparable to those of the $d$-wave phase on the regular lattice and persists across a broad doping range.

This restructuring can be understood in terms of the bond-level disruption introduced by the carpet construction. As discussed in App.\ref{AppendixA}, the systematic removal of sites breaks the local four-bond motifs that support a coherent $d_{x^2-y^2}$ sign alternation between perpendicular directions. On many sites, the coordination shell is incomplete, and the local orthogonality between extended $s$ and $d$ channels is lost. As a result, the self-consistent solution reorganizes into a state in which channel mixing occurs at the local level alternating the global sign structure of the condensate.

Interestingly, while our previous work showed that $T_c$ enhancement for purely on-site $s$-wave superconductivity required finitely ramified fractals such as the Sierpiński gasket, here we observe a substantial increase of the extended $s$-wave superconducting $T_c$ even on the infinitely ramified carpet. This indicates that the distinction between finite and infinite ramification depends sensitively on the symmetry and spatial structure of the pairing channel, and that the interplay between order-parameter symmetry and fractal topology is more subtle than previously anticipated.

Only if $s$-wave component is strictly not allowed, and pure $d$-wave symmetry is imposed on the level of initial seed, the BdG scheme converges to a profile that resembles extended $s$-wave on incomplete local coordination stars, while the global structure of the order parameter retains a sign-changing pattern across the lattice, App. \ref{AppendixB}. In particular, positive and negative bond amplitudes are related by global symmetry operations of the underlying square lattice rather than by local $90^\circ$ rotations of an intact four-bond “cross”. In this sense, the superconducting state on the carpet is a geometrically modified solution whose global transformation properties remain consistent with a $d$-wave representation, even though its local structure is reshaped by bond removal.


  \begin{figure*}[t] 
        \centering
         \includegraphics[width=\textwidth]{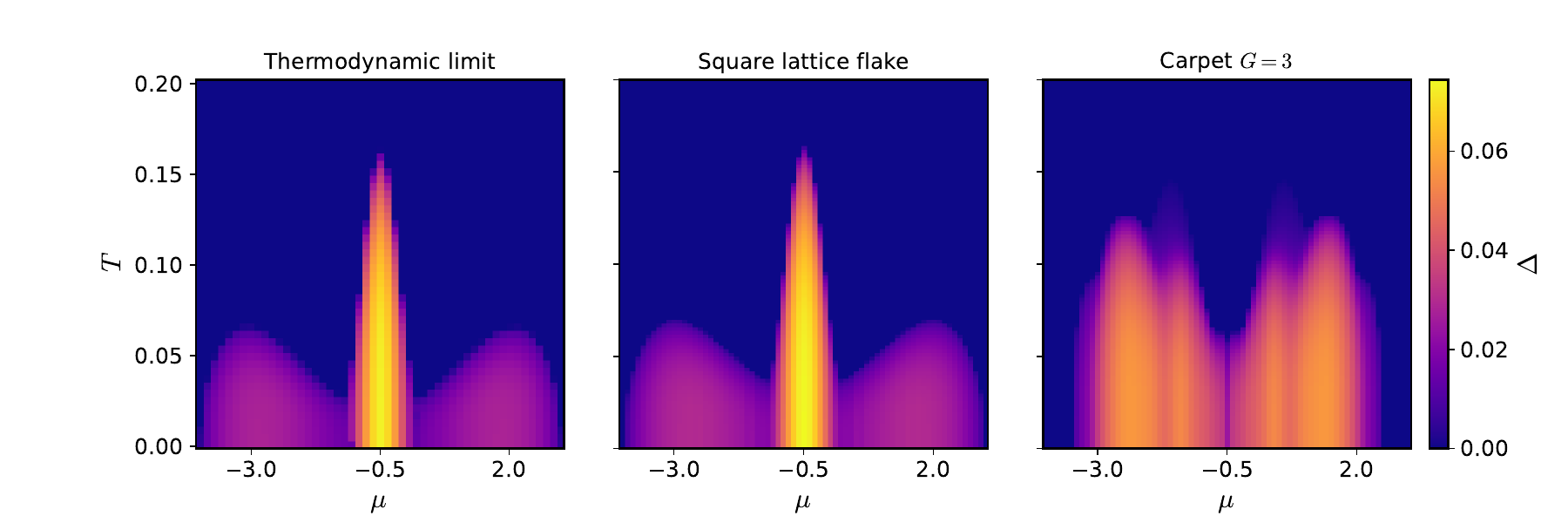}
         \caption{Phase diagrams of the thermodynamic limit of the square lattice at the left, a square lattice flake with side $N_{x}=N_{y}=27$ at the center, and the $G=3$ Carpet at the right.}
         \label{fig:square-carpet}
    \end{figure*}
    \subsection{Triangular Sierpinksi gasket}
    We now turn to the triangular Sierpiński gasket and analyze how the fractal geometry modifies the phase diagram in the presence of extended $s$-wave and $d$-wave pairing channels, Fig. 3.

On the regular triangular lattice, the nearest-neighbor $d$-wave channel spans a two-dimensional irreducible representation, allowing for both real $d$-wave and chiral $d+id$ states. Extended $s$-wave competes with this channel but does not trivially dominate. The superconducting instability is therefore controlled by the interplay between a fully symmetric component and a two-dimensional $d$-subspace.

Upon constructing the Sierpiński gasket, two out of six bonds are removed at many lattice sites, leading to locally incomplete coordination shells. Like in the previous cases, such bond removal generically breaks the orthogonality between extended $s$ and $d$ components at the level of a single coordination star and modifies their relative spectral weights. Importantly, however, the $d$ sector forms a two-dimensional irreducible representation of the lattice symmetry group, corresponding to the $d_{x^2-y^2}$ and $d_{xy}$ channels. Bond removal therefore induces mixing between these channels rather than eliminating the $d$ subspace altogether.

The resulting self-consistent solutions on the gasket do not realize a pure $d$ or pure $d+id$ phase. Instead, the stable superconducting state acquires a mixed character, combining extended $s$ with both $d$ components. In practice, the order parameter develops a finite complex structure, reflecting the coexistence of real and imaginary $d$ components together with a symmetric contribution.

The phase diagram exhibits a clear enhancement of superconductivity relative to the regular lattice. The critical temperature increases across a broad range of chemical potential, and the magnitude of the gap is correspondingly larger. At the same time, the superconducting domes become slightly narrower in chemical potential compared to the regular triangular lattice, although this reduction is modest. This behavior contrasts with our earlier results for purely on-site $s$-wave pairing, where the dome narrowing on fractal geometries was substantially stronger.

Overall, the triangular Sierpiński gasket does not suppress superconductivity; rather, it reshapes the competition between pairing channels, promoting a mixed $s+d+id$ state with enhanced gap amplitude and elevated critical temperature over a somewhat reduced doping window.
    \begin{figure*}[t] 
        \centering
         \includegraphics[width=\textwidth]{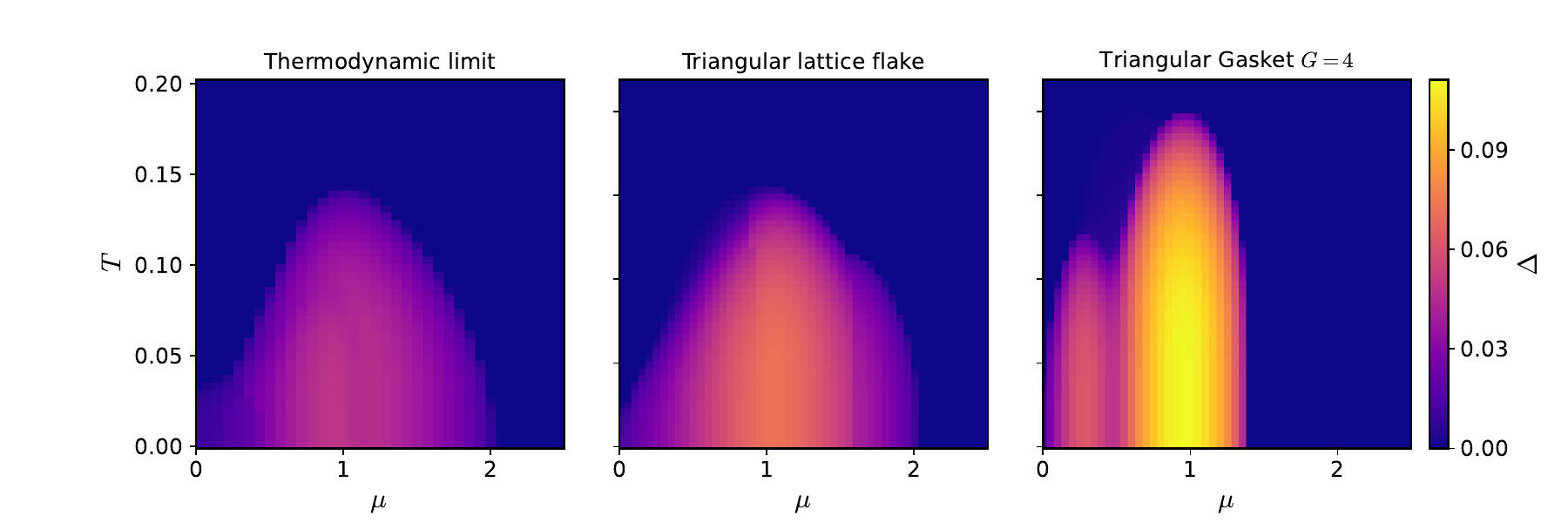}
         \caption{Phase diagrams of the thermodynamic limit of the triangular lattice at the left, an equilateral triangle flake with triangular lattice base with side $N_{x}=33$ at the center, and the $G=4$ Sierpiński gasket at the right.}
         \label{fig:square-carpet}
    \end{figure*}
    \begin{figure*}[t] 
        \centering
         \includegraphics[width=\textwidth]{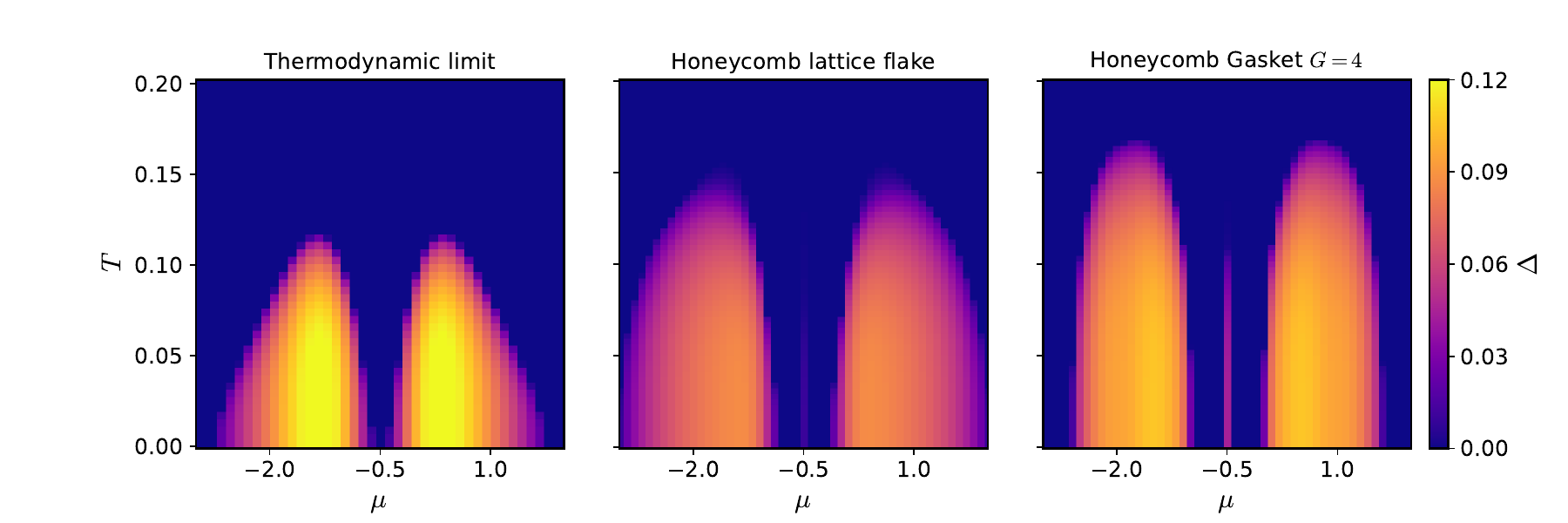}
         \caption{Phase diagrams of the thermodynamic limit of the honeycomb lattice at the left, an equilateral triangle flake with honeycomb base at the center, and the $G=4$ Sierpiński gasket at the right.}
         \label{fig:square-carpet}
    \end{figure*}
    \subsection{Hexagonal Sierpinksi gasket}
    We finally consider the Sierpiński gasket constructed from the honeycomb lattice. In contrast to the square and triangular cases, the superconducting behavior here remains structurally simple.

On the regular honeycomb lattice, the phase diagram exhibits two symmetric superconducting domes centered around a characteristic value of the chemical potential (approximately $\mu \simeq -0.5$ for the parameters considered), Fig.4. The symmetry of these domes reflects the bipartite nature of the lattice and the associated particle–hole structure of the spectrum.

Upon constructing the Sierpiński gasket, this symmetry is preserved: the superconducting domes remain centered at the same chemical potential, and no noticeable shift or asymmetry is introduced by the fractal geometry. This robustness is consistent with the underlying bipartite structure, which remains intact despite the removal of sites.

The primary effect of fractalization in this case is a quantitative enhancement of superconductivity. Both the critical temperature $T_c$ and the zero-temperature gap magnitude increase relative to the regular lattice. Unlike the square and triangular geometries, we do not observe a nontrivial competition between pairing channels or the emergence of mixed states. The superconducting order parameter retains a uniform $s$-wave character across the lattice, without additional sign structure or complex components.

Thus, on the honeycomb-derived Sierpiński gasket, fractal geometry enhances superconductivity while preserving the symmetry and qualitative structure of the pairing state.

\subsection{Phase stiffness}

To quantify the degree of global phase coherence of the self-consistent superconducting states, we compute the phase stiffness by applying a weak electric field along the $x$ axis and extracting the resulting current response. The spatial maps in Fig.~\ref{fig:stiffness} show that the stiffness remains considerable in all cases, indicating that the condensate can sustain a coherent response across the entire system. At the same time, the stiffness is not equally uniform in space for the different geometries. For the hexagonal gasket we find a comparatively homogeneous stiffness profile. By contrast, for the Sierpiński carpet and the triangular gasket the stiffness is more spatially structured, with clear variations across the sample.

Although the self-consistent pairing amplitude is finite and the computed phase stiffness remains sizable for all geometries considered, this does not by itself guarantee macroscopic phase coherence. Establishing true long-range or quasi-long-range order requires mapping the low-energy sector onto an effective Josephson $XY$ model \cite{HarlandXY} and analyzing vortex excitations beyond mean-field theory. For the Sierpiński gasket in the thermodynamic limit, it is known that the vortex self-energy remains finite and no Berezinskii–Kosterlitz–Thouless transition occurs; correlations decay exponentially at any finite temperature. In that strict infinite-fractal limit, a nonzero mean-field stiffness therefore does not imply algebraic order \cite{XYSierpinski}.

However, our focus is on fractal structures of finite generation, which are the physically realizable systems in nanofabrication experiments. In such systems the hierarchical geometry introduces a finite maximal length scale, cutting off the infrared proliferation of vortices that destroys quasi-long-range order in the infinite gasket. As a result, the effective phase dynamics differs qualitatively from the thermodynamic-limit scenario. Ongoing numerical analysis of the corresponding classical $XY$ model on finite-depth gaskets indicates that phase coherence can persist up to finite temperatures set by the largest structural scale. Furthermore, at strictly zero temperature the relevant problem is that of the quantum $XY$ model. Preliminary results show that quantum fluctuations do not eliminate algebraic order in the ground state for finite ramification depth. This suggests that the enhanced pairing amplitudes found at the mean-field level can translate into physically meaningful superconducting states once realistic system size and dimensional crossover effects are taken into account. A detailed analysis of these phase-fluctuation effects will be presented in a separate work.



\begin{figure*}[t]
    \centering
    \includegraphics[width=\textwidth]{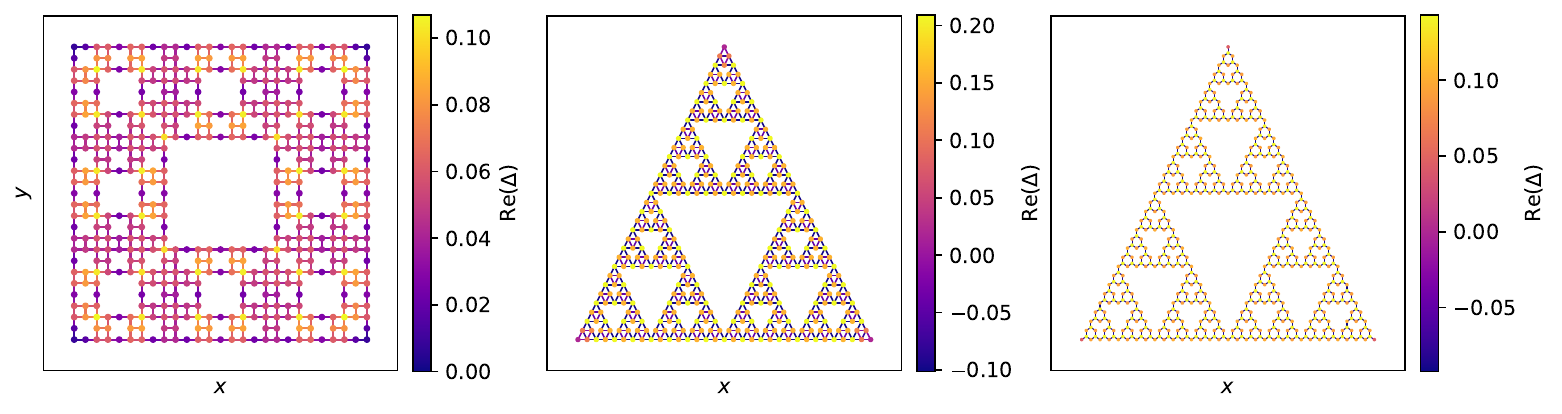}
    \caption{Profiles of the order parameter for the different geometries/pairing states at $T=0.001$ and $\mu = -2.7$, $\mu = 1.15$, and $\mu = 1$ shown in panels (a)--(c) respectively. Circular markers shows on-site superconductivity and the edge color corresponds to the extended component.}\textbf{}
    \label{fig:delta}
\end{figure*}
\begin{figure*}[t]
    \centering
    \includegraphics[width=\textwidth]{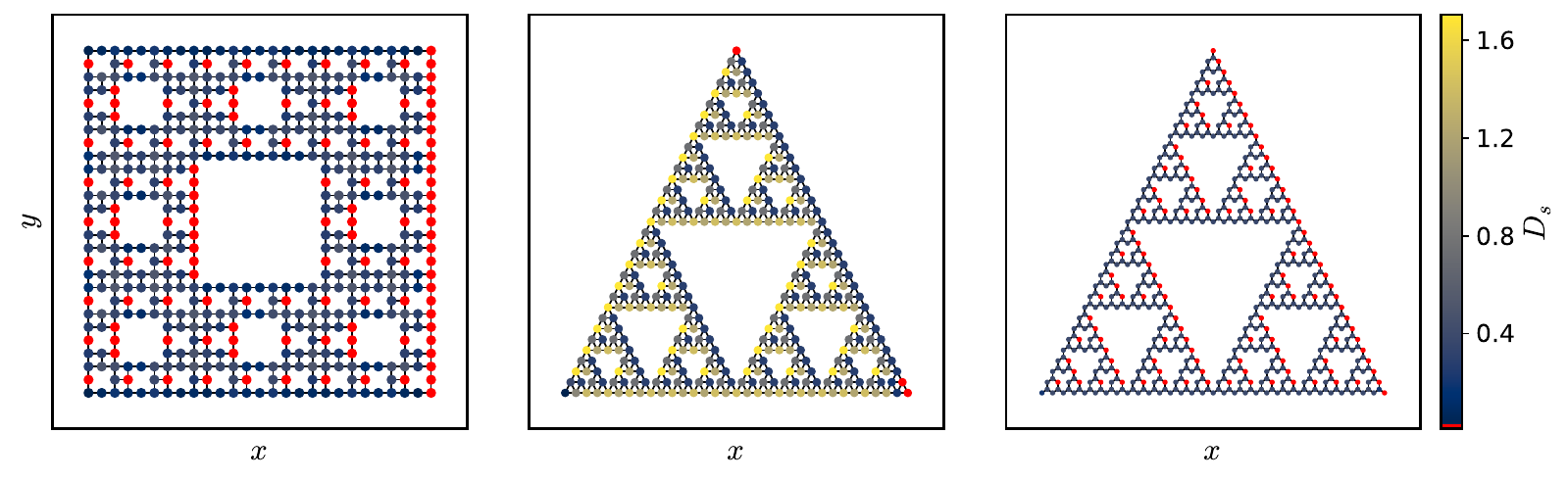}
    \caption{Phase stiffness for the different geometries/pairing states at $T=0.001$ and $\mu = -2.7$, $\mu = 1.15$, and $\mu = 1$ shown in panels (a)--(c) respectively.}
    \label{fig:stiffness}
\end{figure*}

	\section{Discussion}
In this work, we studied how fractal geometry affects superconducting order parameters with different symmetries in the extended Hubbard model. Our results show that the stability of pairing channels depends sensitively on how the spatial structure of Cooper pairs adapts to the geometric constraints imposed by the lattice. Fractalization both renormalizes transition temperatures and reorganizes the internal structure of the superconducting state.

The contrast between the regular square lattice and the Sierpiński carpet illustrates this most clearly. On the regular square lattice with nearest-neighbor attraction, $d_{x^2-y^2}$ pairing dominates, forming a broad superconducting dome with high critical temperature, while extended $s$-wave appears only as weak satellite domes. On the Sierpiński carpet, however, the pure $d$-wave phase is no longer stabilized as an isolated solution in the corresponding parameter region as long as $s$-wave is in principle allowed. Instead, the self-consistent state transforms into an extended-$s$ component at the level of local coordination shells, and its critical temperature increases relative to the regular lattice case. 
In this sense, the carpet does not simply suppress superconductivity but rather reorganizes the dominant pairing channel.

This restructuring can be traced to the bond-level disruption introduced by the carpet construction. Removing sites breaks the local orthogonality between extended-$s$ and $d$ components, leading to channel mixing on incomplete coordination stars. The superconducting solution therefore reflects a compromise between local geometric constraints and global symmetry requirements. Extended-$s$–like components, being less sensitive to the presence of perpendicular bond pairs, gain relative weight in the fractal geometry.

For the triangular lattice and its Sierpiński gasket counterpart, the situation is qualitatively different. On the regular triangular lattice, nearest-neighbor $d$-wave pairing spans a two-dimensional irreducible representation, allowing both real $d$ and chiral $d+id$ states. On the gasket, bond removal again induces local mixing between extended-$s$ and $d$ components, but the two-dimensional $d$-subspace remains active. The stable superconducting state acquires a mixed $s+d+id$ character rather than collapsing into a purely symmetric solution. We observe an enhancement of both the critical temperature and the gap amplitude relative to the regular triangular lattice, while the superconducting domes narrow only modestly in chemical potential. In this case, fractal geometry reshapes the balance between pairing channels without eliminating the underlying $d$-wave structure.

The honeycomb-derived Sierpiński gasket presents a simpler scenario. On the regular honeycomb lattice, superconductivity forms two symmetric domes centered around a characteristic chemical potential, consistent with the bipartite structure of the lattice. This symmetry is preserved upon fractalization: the domes remain centered at the same chemical potential, with no significant shift or asymmetry. The primary effect of the gasket geometry is quantitative rather than qualitative: both the critical temperature and the gap magnitude increase, while the pairing symmetry remains purely $s$-wave. In this case, fractal geometry enhances superconductivity without fundamentally altering the structure of the order parameter.

Taken together, these results show that the distinction between finite and infinite ramification, which proved central in our earlier study of purely on-site $s$-wave pairing, becomes more nuanced when anisotropic channels are included. The infinitely ramified Sierpiński carpet, which did not enhance pure on-site $s$-wave pairing in isolation, significantly enhances extended-$s$–dominated states when competing against $d$-wave order. Conversely, the finitely ramified triangular gasket supports enhancement of a mixed $s+d+id$ phase. Fractal geometry therefore acts as a symmetry-selective environment: rather than uniformly boosting or suppressing superconductivity, it reshapes the competition between pairing channels in a symmetry-dependent manner.

Our calculations also show that the superfluid stiffness remains finite throughout the superconducting regions of all considered geometries. However, a finite mean-field stiffness does not automatically imply true long-range order. According to the Mermin–Wagner theorem, continuous symmetries cannot be spontaneously broken at finite temperature in dimensions $d \le 2$. Fractal lattices are characterized by non-integer Hausdorff dimensions, $d_H \simeq 1.58$ for the Sierpiński gasket and $d_H \simeq 1.89$ for the Sierpiński carpet, both below two. In principle, this places them in a regime where phase fluctuations may strongly suppress or even eliminate finite-temperature long-range order in the thermodynamic limit.

At the same time, the mechanisms responsible for the enhancement of the pairing amplitude such as modified density of states, boundary-dominated spectra, and suppression or reshaping of competing channels are distinct from those governing phase coherence. Two additional considerations are therefore important. First, our analysis focuses on finite-generation fractals of experimentally relevant size, where macroscopic coherence can persist at finite temperature. Second, mapping the low-energy sector onto an effective Josephson $XY$ model suggests that finite-depth structures can sustain quasi-long-range order, while at strictly zero temperature quantum fluctuations in the corresponding quantum $XY$ model do not destroy algebraic order. Thus, although phase fluctuations remain a central open problem beyond mean-field theory, the geometric enhancement of pairing amplitudes we observe is not trivially nullified by dimensional arguments alone.

Several directions for future work naturally follow. To address phase fluctuations and correlation effects beyond mean field, more advanced methods are required. Real-space extensions of the GW \cite{GW} or constrained random phase approximation \cite{cRPA} could incorporate screening and vertex corrections while respecting fractal geometry. Neural quantum states \cite{NQS} and modern variational Monte Carlo techniques offer another promising route for treating both strong correlations and irregular lattices. It would also be valuable to investigate the repulsive Hubbard model on fractal geometries, where pairing emerges from exchange processes rather than direct attraction, and the interplay with lattice topology may differ qualitatively.

In our recent work on hyperbolic lattices \cite{Bashmakov}, we demonstrated that modifying boundary geometry and lattice symmetries can dramatically enhance superconductivity by increasing the number of low-energy states. Fractals share with hyperbolic lattices the property that boundary effects dominate over bulk in the thermodynamic limit. The highly structured boundaries of Sierpiński geometries may therefore provide additional avenues for tuning the low-energy spectrum. Systematic exploration of boundary terminations, defect patterns, or hybrid structures combining fractal and crystalline motifs could reveal further routes toward optimizing $T_c$.

Experimentally, two main approaches appear promising. Scanning tunneling microscopy techniques \cite{Khajetoorians}, which have already been used to assemble fractal structures atom by atom \cite{STM}, could directly probe the spatial profile of the superconducting gap and test the predicted competition between pairing channels. Alternatively, high-resolution lithography \cite{lithography} may enable fabrication of fractal patterns in thin superconducting films, allowing transport and thermodynamic measurements of phase diagrams on engineered geometries.

More broadly, we have focused here on Sierpiński structures, but the space of fractal geometries is vast. An especially interesting direction would be to study periodic lattices constructed from finite-depth fractal supercells \cite{fractal_crystal}. Such systems would interpolate between fully fractal and conventional crystalline structures, allowing systematic investigation of how much fractality is required to achieve enhancement and whether universal scaling relations connect geometric measures such as Hausdorff dimension, ramification index, or boundary-to-bulk ratio to superconducting observables like $T_c$, gap magnitude, and stiffness.

The central lesson of this work is that geometry and order-parameter symmetry are deeply intertwined. Fractal lattices reshape the symmetry landscape in which pairing develops leading to considerable restructuring of the phase diagram. By selectively stabilizing some channels and reshaping others, lattice geometry itself becomes a design parameter offering a complementary route to engineering superconducting states beyond conventional tuning of interactions and doping.
        
	
\bibliographystyle{apsrev4-1}
\bibliography{fractal_sc_fixed}
\appendix
\section{Local channel mixing induced by bond removal}
\label{AppendixA}
In this Appendix we illustrate, at the level of a single coordination shell, how bond removal modifies the structure of nearest-neighbor pairing channels. The purpose is purely algebraic: to show explicitly how extended $s$-wave and $d$-wave components cease to be orthogonal once the local bond environment is incomplete.

We consider the ``star'' of nearest-neighbor bonds emanating from a given site. Restricting the order parameter to these bonds, we represent it as a vector whose components correspond to bond directions. For a given set of available directions $\mathcal S$, and a set of channel form factors ${\phi_\alpha(\theta)}$ evaluated on these directions, we define the local Gram matrix
\begin{equation}
    M_{\alpha\beta}(\mathcal S)
= \sum_{\theta_n\in\mathcal S}
\phi_\alpha(\theta_n)\phi_\beta(\theta_n),
\end{equation}
where $\alpha,\beta$ label pairing channels. Orthogonality of channels corresponds to $M_{\alpha\beta}=0$ for $\alpha\neq\beta$. Once bonds are removed, this orthogonality may be lost.

For the square lattice, the four nearest-neighbor directions are
\begin{equation}
    \theta \in \left\{0,\frac{\pi}{2},\pi,\frac{3\pi}{2}\right\}.
\end{equation}
We define the extended $s$-wave and $d_{x^2-y^2}$ channel vectors as
\begin{gather}
    \phi_s(\theta)=1, \\
    \phi_d(\theta)=
\begin{cases}
+1,& \theta=0,\pi,\\
-1,& \theta=\frac{\pi}{2},\frac{3\pi}{2}.
\end{cases}
\end{gather}
On the complete star $\mathcal S_0$ of four bonds, the Gram matrix in the basis $(s,d)$ is
\begin{equation}
    M^{(\square)}(\mathcal S_0)
=
\begin{pmatrix}
4 & 0 \\
0 & 4
\end{pmatrix}.
\end{equation}
Now remove one bond, for instance, the direction $\theta=\pi/2$. The remaining set is
\begin{equation}
   \mathcal S=\left\{0,\pi,\frac{3\pi}{2}\right\}.
\end{equation}
Restricting the channel vectors to $\mathcal S$ gives
\begin{equation}
    \boldsymbol{\phi}_s=(1,1,1), \qquad
\boldsymbol{\phi}_d=(1,1,-1).
\end{equation}
The corresponding Gram matrix becomes
\begin{equation}
    M^{(\square)}(\mathcal S)
=
\begin{pmatrix}
3 & 1 \\
1 & 3
\end{pmatrix}.
\end{equation}
The off-diagonal element $M_{sd}=1$ is nonzero, demonstrating that extended $s$-wave and $d$-wave components are no longer orthogonal once a single bond is removed, and the two channels mix.

Similarly, for the triangular lattice, the six nearest-neighbor directions are
\begin{equation}
    \theta \in
\left\{
0,\frac{\pi}{3},\frac{2\pi}{3},
\pi,\frac{4\pi}{3},\frac{5\pi}{3}
\right\}.
\end{equation}
We define the extended $s$-wave channel
\begin{equation}
    \phi_s(\theta)=1,
\end{equation}
and a two-dimensional $d$-wave basis
\begin{equation}
    \phi_1(\theta)=\cos(2\theta), \qquad
\phi_2(\theta)=\sin(2\theta).
\end{equation}
On the complete six-bond star $\mathcal S_0$, the Gram matrix in the basis $(s,1,2)$ is
\begin{equation}
    M^{(\triangle)}(\mathcal S_0)
=
\begin{pmatrix}
6 & 0 & 0 \\
0 & 3 & 0 \\
0 & 0 & 3
\end{pmatrix}.
\end{equation}
Now remove two consecutive bonds (which is typical for the gasket cut), for instance $\theta=\pi/3$ and $2\pi/3$. The remaining set is
\begin{equation}
    \mathcal S=
\left\{
0,\pi,\frac{4\pi}{3},\frac{5\pi}{3}
\right\}.
\end{equation}
Evaluating the basis functions on this set yields
\begin{gather}
    \boldsymbol{\phi}_s=(1,1,1,1), \\
    \boldsymbol{\phi}_1=\left(1,1,-\frac12,-\frac12\right), \\
    \boldsymbol{\phi}_2=\left(0,0,\frac{\sqrt3}{2},-\frac{\sqrt3}{2}\right).
\end{gather}
The Gram matrix becomes
\begin{equation}
    M^{(\triangle)}(\mathcal S)
=
\begin{pmatrix}
4 & 1 & 0 \\
1 & \frac{5}{2} & 0 \\
0 & 0 & \frac{3}{2}
\end{pmatrix},
\end{equation}
again, with off-diagonal components mixing the $s$ and $d$ channels.

These local examples explicitly demonstrate that bond removal generically induces mixing between pairing channels and modifies their relative spectral weights, even before solving the full self-consistent problem.
\section{Pure d-wave on higher-generation Sierpinski carpet}
\label{AppendixB}
In this Appendix, we show spatial profile of $\Delta$ on larger-scale Sierpinski carpet obtained by mean of KPM in the case when only $d$-wave component is present. Fig. 7, 8, and 9 show how the result depends on the number of retained moments. One can note that upon convergence the profile itself acquires a non-trivial fractal pattern with alternating regions of positive and negative $\Delta$ emerging at diverse scales obeying a $d$-wave symmetry globally, but not on the level of local coordination shells.

Fig.~\ref{fig:delta_kpm_4_gasket_r} and~\ref{fig:delta_kpm_4_gasket_i} show the real and imaginary parts of the converged order parameter for a $G=7$ triangular Sierpinski gasket. Although initially seeded with a $d_{x^2-y^2}+id_{xy}$ order, the converged result does not resemble the triangular $d+id$ symmetry. When taking the absolute value, it shows a homogeneous profile of $\Delta$. 

\begin{figure*}
    \centering
    \includegraphics[width=\textwidth]{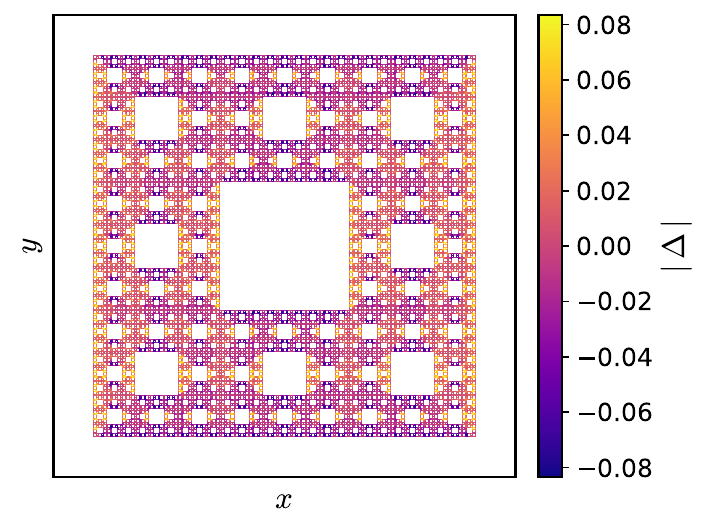}
    \caption{$d_{x^2-y^2}$ wave order parameter profile for $G=5$ carpet converged using the KPM and retaining 200 moments at $T=0$ and $\mu=-2.8.$}
    \label{fig:delta_kpm_5(1)}
\end{figure*}

\begin{figure*}
    \centering
    \includegraphics[width=\textwidth]{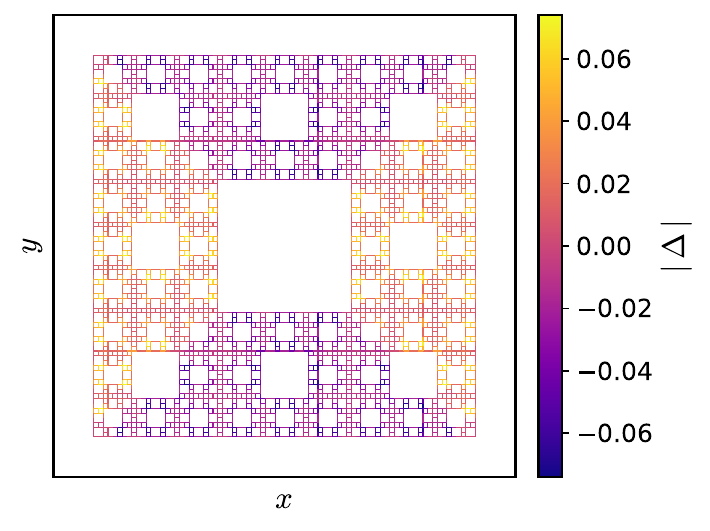}
    \caption{$d_{x^2-y^2}$ wave order parameter profile for $G=4$ carpet converged using the KPM and retaining 1000 moments at $T=0$ and $\mu=-2.8.$}
    \label{fig:delta_kpm_4}
\end{figure*}

\begin{figure*}
    \centering
    \includegraphics[width=\textwidth]{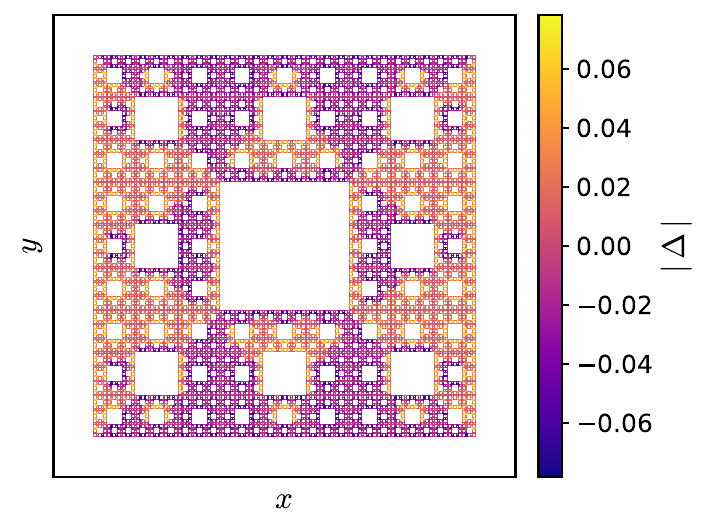}
    \caption{$d_{x^2-y^2}$ wave order parameter profile for $G=5$ carpet converged using the KPM and retaining 2000 moments at $T=0$ and $\mu=-2.8.$}
    \label{fig:delta_kpm_5(2)}
\end{figure*}

\begin{figure*}
    \centering
    \includegraphics[width=\textwidth]{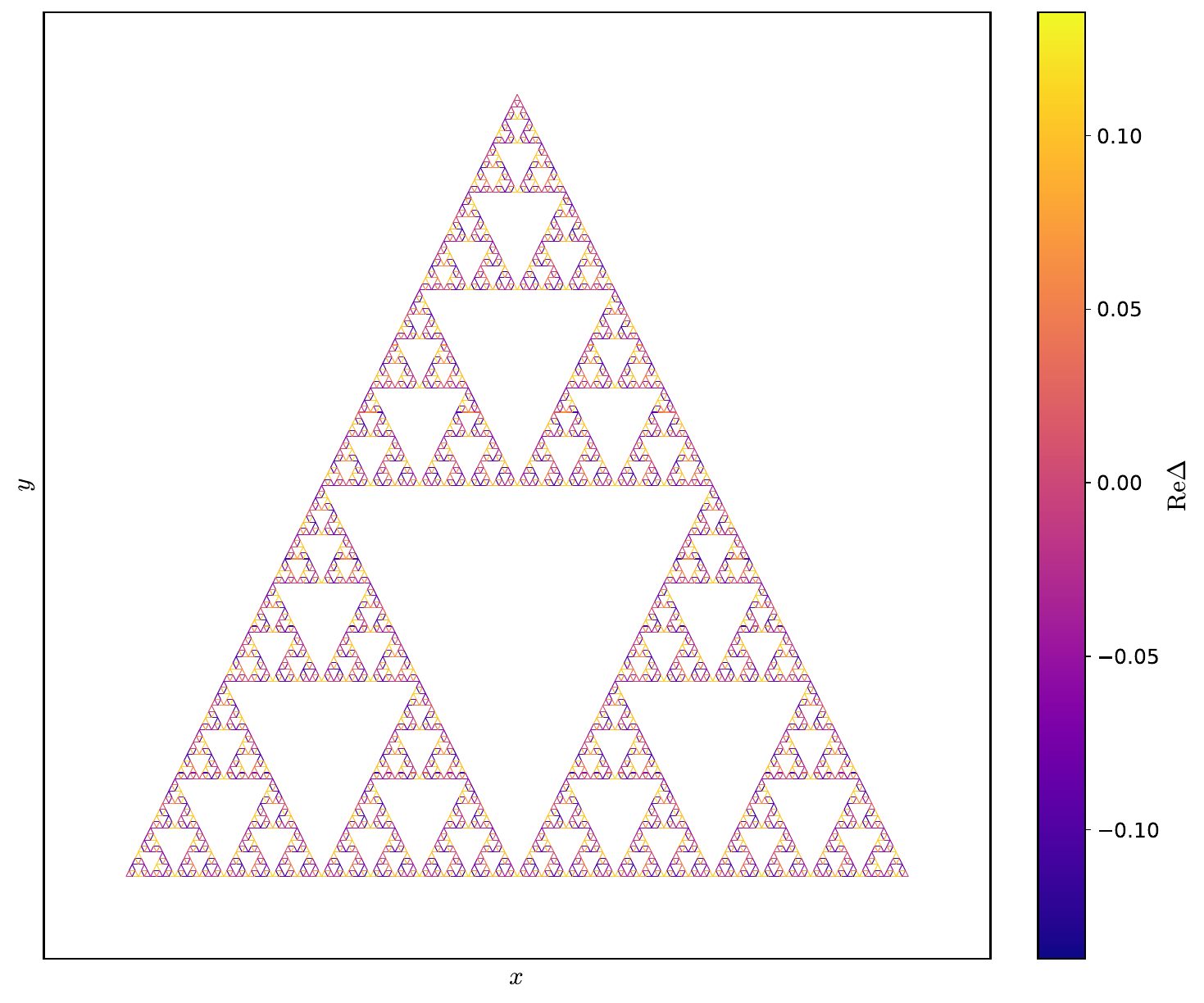}
    \caption{$d_{x^2-y^2}+id_{xy}$ wave order parameter profile (real part) for $G=7$ carpet converged using the KPM and retaining 1000 moments at $T=0$ and $\mu=1.15.$}
    \label{fig:delta_kpm_4_gasket_r}
\end{figure*}

\begin{figure*}
    \centering
    \includegraphics[width=\textwidth]{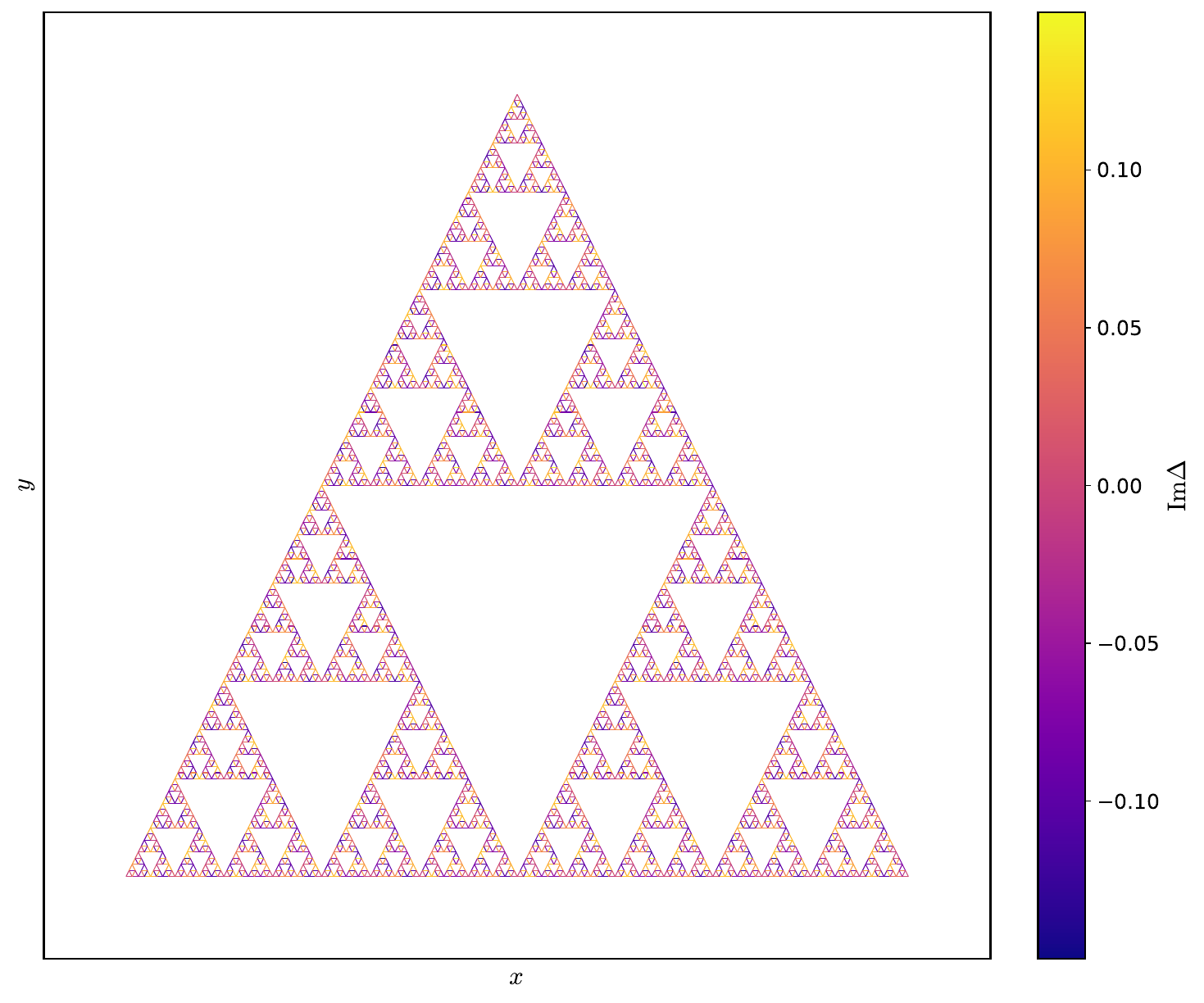}
    \caption{$d_{x^2-y^2}+id_{xy}$ wave order parameter profile (imaginary part) for $G=7$ carpet converged using the KPM and retaining 1000 moments at $T=0$ and $\mu=1.15.$}
    \label{fig:delta_kpm_4_gasket_i}
\end{figure*}

\end{document}